\documentclass[11pt]{article}
\usepackage[numbers]{natbib}
\usepackage{xurl}
\usepackage{hyperref}
\usepackage{breakurl}

\usepackage{authblk}

\author[1]{Murali Mani}
\author[2]{Jie Shen}
\author[3]{Tejaswi Manchineella}
\author[4]{Ira Woodring}
\author[5]{Jing Bai}
\author[6]{Robert Benard}
\author[7]{E Shirl Donaldson}

\affil[1]{University of Michigan, Flint\\ \texttt{mmani@umich.edu}}
\affil[2]{University of Michigan, Dearborn\\ \texttt{shen@umich.edu}}
\affil[3]{Grand Valley State University\\ \texttt{manchite@gvsu.edu}}
\affil[4]{Grand Valley State University\\ \texttt{woodriir@gvsu.edu}}
\affil[5]{Washtenaw Community College\\ \texttt{jswanson@wccnet.edu}}
\affil[6]{Mott Community College\\ \texttt{robert.benard@mcc.edu}}
\affil[7]{University of Michigan, Flint\\ \texttt{shirld@umich.edu}}

\title{We Need to Effectively Integrate Computing Skills Across Discipline Curricula}

\begin{document}
\sloppy

\maketitle

Computing is increasingly central to innovation across a wide range of disciplinary and interdisciplinary problem domains~\cite{NSFIUSECFP}. Students across non-computing disciplines need to apply sophisticated computational skills and methods to fields as diverse as biology, linguistics, and art. Furthermore, computing plays a critical role in “momentous geopolitical events,” such as elections in several countries including the US, and is changing how people “work, collaborate, communicate, shop, eat, travel, get news and entertainment, and quite simply live”~\cite{GP2018}. Traditional computing courses, however, fail to equip non-computing discipline students with the necessary computing skills – if they can even get into classes packed with CS majors. A pressing question facing academics today is: \emph{How do we effectively integrate computing skills that are useful for the discipline into discipline curricula?}

One approach is CS+X courses, which are themed or contextualized computing courses typically with different learning goals than traditional computing courses. The goal of such courses is to prepare students in discipline X with the computing skills necessary for X, and to provide opportunities to explore the connections between CS and X. Examples of such themes that different computing educators have studied include media computation, biology, bioinformatics, scientific computing, scientific discovery, expression in language and arts, social justice, law, data analysis, data visualization, web programming, game development, robotics, and AI.

While CS+X courses have proven useful in several scenarios, they often have limitations in meeting the necessary computing-related learning outcomes of discipline X. Typically, discipline X will have practical limitations on the number of CS+X courses that can be integrated into their curriculum. How do we effectively sequence such CS+X courses in the curriculum? If introduced too early, students may not form effective connections between future concepts they will learn in discipline X and computing. If introduced too late, students might not be interested in re-learning their X concepts using computing, and hence will not be able to use computing skills effectively for their work in X. Furthermore, restricting computing to CS+X courses is contrary to the pervasiveness of computing and will likely result in computing not being a learning goal for most courses in discipline X.

We propose an alternative to CS+X courses. We advocate that courses in discipline X include the computing relevant to the learning outcomes of that course, as used by practitioners in X. We refer to the computing skills relevant to a course in discipline X as an \textbf{``ounce of computing skills,”} to highlight our belief regarding the amount of computing to be integrated in that course, and to contrast with teaspoon programming languages~\cite{GDN+2023}. In this article, we outline our insights regarding the development of an ounce of computing skills for a discipline course, and the evaluation of the developed ounce. The key takeaways are that the goal has to be to advance students in their disciplines, and only the disciplinary experts can tell us how computing is used in that discipline. Computer scientists know how to teach computing, but the classes can’t be about CS values. The disciplinary values are paramount.

\section*{Developing An Ounce of Computing Skills}

Discipline instructors know how computing is used in their discipline and understand the discipline’s pedagogical practices. Computing experts, who are passionate about computing education in various contexts, can teach computing skills tailored to specific learning goals. Students from the disciplines, who will ultimately use these ounces for learning, are important as they can provide insights into what is motivating and engaging from the student’s perspective~\cite{SRAD1997}. Therefore, developing an ounce of computing skills requires discipline instructors and computing experts to act as co-designers~\cite{SS2008}, with discipline students serving as design informants~\cite{SRAD1997}. These developed ounces could take multiple formats as appropriate; some options from the literature include customized educational technology tools, instructional videos, and focused minimal manuals~\cite{CSFM1987}.

The delivery and integration of an ounce of computing skills in a discipline course must be led by the discipline instructor. They know their students, including their preparation, the discipline’s pedagogical practices, and the previous concepts the students have learned. They can help students form appropriate connections and achieve effective transfer. However, discipline instructors may often need additional support for the integration of the ounce, which computing experts should be ready to provide. This support is typically scaffolding that fades over time, though longer-term support may be needed in some scenarios, requiring additional resources. Integration methods may include traditional lectures, (virtual) labs, group activities, flipped classroom approaches, projects, and other interactive learning experiences. The integration of the developed ounce should be guided by best practices in equity to ensure the ounces are accessible to all students, as well documented in computing education~\cite{Kapor2021, AIICEBarr}. An example of an ounce of computing developed at UM-Flint can be found at \url{https://umresearcher.github.io/computingDisciplines/HealthSciences/calculators.html}. Here, we leveraged developments in software development processes, including GenAI-assisted-coding, to create tools (requiring only minimal effort for software development) aimed at alleviating math anxiety and low self-efficacy commonly observed in students in epidemiology courses, thereby ensuring increased accessibility of epidemiology concepts.

\section*{Evaluating An Ounce of Computing Skills}

An ounce of computing skills should be evaluated against five goals: effectiveness, usefulness, usability, equity, and sustainability (see Table~\ref{tab:goals}). These goals are inspired by efforts such as~\cite{RWI2010}, which outlines desired characteristics of computational thinking (CT) tools in K-12 settings. Here, the authors specify goals of low threshold, high ceiling, scaffolding, enabling transfer, equity, systemic adoption and sustainability. Compared to~\cite{RWI2010}, we introduce the goal of effectiveness to ensure that our ounce effectively meets the necessary learning outcomes. Usefulness ensures that our ounce includes relevant computing skills and maintains an appropriate cognitive load. Our goal of usability aligns with the goal of low threshold in~\cite{RWI2010}. While goals of high ceiling, scaffolding, and enabling transfer may not be universally applicable for all our ounces, they can be incorporated as needed under our effectiveness goal. The goal of systemic adoption, essential for~\cite{RWI2010} to facilitate widespread adoption of the CT tool in schools and districts, is not applicable for our ounces.

Let us define the goal of sustainability, which is not considered in~\cite{RWI2010}. A sustainable ounce of computing skills should have the following features:

\begin{itemize}

\item The discipline instructor should be able to integrate the ounce with little assistance. 

\item The discipline instructor should be able to integrate the ounce with a level of effort comparable to that typically required for other discipline-specific course learning outcomes.

\item The discipline instructor should be able to update and maintain the ounce with little assistance, and with a level of effort comparable to that typically required for updating and maintaining other discipline course materials.

\end{itemize}

\begin{table*}[ht]
  \caption{Goals for Our Ounces of Computing Skills and Their Meanings}
  \label{tab:goals}
  \begin{tabular}{p{1in}p{3.5in}}
    \hline
    \textbf{Goal} & \textbf{Meaning} \\
    \hline
    Effectiveness & The ounces help achieve student learning outcomes. \\
    Usefulness & The computing skills are relevant and applicable to the discipline. \\
    Usability & The students can easily use the ounces. \\
    Equity & All students, regardless of background, benefit equally from the ounces. \\
    Sustainability & The discipline instructor can sustainably integrate the ounces into their courses, as well as update and maintain the ounces over time. \\
    \hline
\end{tabular}
\end{table*}

A discipline instructor may need additional support for the integration of the ounce as well as for the update and maintenance of the ounce, which may be provided by the computing experts. Such support could be a scaffolding support that fades over time, or could be a longer-term support that needs additional resources. While any support needed by the discipline instructor for the integration of an ounce could fade over time, the discipline instructor is likely to need periodic support (say, every few years) for the update and maintenance of the ounce. The sustainability of an ounce of computing skills, including the effectiveness and the fading of any scaffolding support, must be evaluated.

\section*{Conclusions}

We have outlined an approach that will fundamentally transform the integration of computing skills into the education of students from various disciplines by embedding discipline-specific computing skills into individual discipline courses. This will empower students from non-computing disciplines to form effective connections between computing and their field, significantly enhancing their professional capabilities, while expanding the utility of computing in a wide range of professional contexts. Our preliminary research indicates a high level of interest from non-computing discipline instructors across institutions and teaching modalities (in addition to the ounce developed for epidemiology courses, we have begun developing ounces of computing for a management course, for a discrete math course, and for a health finance course), indicating that our approach is both feasible and likely to be widely adopted.

Furthermore, this project will expose a diverse set of students, including women and individuals from groups underrepresented in computing, to computing from the perspective of their individual disciplines and interests, which has been an elusive goal for the computing community for several decades.

We hope this article will inspire educators to explore our proposed approach, as well as alternatives, for integrating discipline-specific computing skills into discipline curricula. This will ensure that computing education is not confined to traditional pathways that often lack diversity but is instead integrated into various disciplines, fostering a richer and more inclusive learning experience. By doing so, we aim to produce a workforce that is not only proficient in computing but also reflective of the diverse society it serves.  Overall, this will democratize computing education, making it accessible and relevant to a broader range of students, and break down the barriers that have traditionally limited access to computing knowledge.

\bibliographystyle{plainnat}

\end{document}